\documentclass[12pt]{article}
\oddsidemargin = \evensidemargin
\usepackage{epsfig}
\usepackage{amssymb}
\usepackage{colordvi}

\begin{document}

\title{\bf Stabilization of \\ Causally and Non--Causally \\ Coupled Map Lattices}

\author{Harald Atmanspacher$^{1,2}$ and Herbert Scheingraber$^1$ \\ \\
1 Center for Interdisciplinary Plasma Science, \\
Max-Planck-Institut f\"ur extraterrest\-rische Physik, \\ 85740 Garching, Germany \smallskip \\
2 Institute for Frontier Areas of Psychology and Mental Health, \\ 
Wilhelmstr.~3a, 79098 Freiburg, Germany}

\date{}
\maketitle

\bigskip
Accepted for publication in {\it Physica A}

\bigskip


\bigskip


\vskip 1cm

\begin{abstract}

Two-dimensional coupled map lattices have global stability properties
that depend on the coupling between individual maps and their neighborhood. 
The action of the neighborhood on individual maps can be implemented in 
terms of ``causal'' coupling (to spatially distant past states) or 
``non-causal'' coupling (to spatially distant simultaneous states). 
In this contribution we show that globally stable behavior of coupled map
lattices is facilitated by causal coupling, thus indicating a surprising 
relationship between stability and causality. The influence of causal 
versus non-causal coupling for synchronous and asynchronous updating 
as a function of coupling strength and 
for different neighborhoods is analyzed in detail. 
 
\end{abstract}


\vfill\eject

\section{Introduction} 

Coupled map lattices (CMLs) are arrays of states whose values are continuous,
usually within the unit interval, over discrete space and time. Starting with
Turing's seminal work on morphogenesis [Turing 1952], they have
been used to study the behavior of complex spatiotemporal systems for
50 years. More recently, Kaneko and collaborators have established many
interesting results for CMLs (cf.~Kaneko [1993], Kaneko and Tsuda [2000]) as generalizations of cellular automata, whose state values are discrete.    

One key motivation for modeling spatiotemporally extended systems with CMLs
is to simplify the standard approach in terms of partial differential
equations. And, of course, CMLs would not have become accessible without
the rapid development of fast computers with large storage capacities.
Within the last decades, CMLs have been applied to the study of areas as diverse
as social systems, ecosystems, neural networks, spin lattices, Josephson junctions,
multimode lasers, hydrodynamical turbulence, and others (cf.~the special 
journal issues {\it CHAOS} {\bf 2}(3), 1992, and {\it Physica D} {\bf 103}, 1997).       

A compact characterization of a CML over two
spatial dimensions with one time parameter is given by:
\begin{equation}
g_{n+1}(x_{ij}) = (1-\epsilon) f_n(x_{ij}) + {\epsilon\over N} \sum_{k=1}^N f_n(x_k)  
\end{equation}
For $f_n(x)$, the iterative logistic map 
characterized by the function $x \mapsto rx(1-x)$
is mostly used, where $r$ is a control parameter with $0 <r \le 4$,
and $n$ represents the time step of the iteration.
The indices $i,j$ are used to label the position of each cell (or site) in the 
lattice. $N$ is the number of cells defining the neighborhood 
of each cell (with periodic boundary conditions),
and $k$ runs over all neighboring cells. The parameter $\epsilon$
specifies the coupling between each cell and its neighborhood (and is usually 
considered as constant over time and space). Thus, the value 
of $g_{n+1}(x_{ij})$ is a convex combination of the value at each individual cell 
and the mean value of its local neighborhood.   

For $\epsilon\rightarrow 0$, there is no coupling at all; hence, local
neighborhoods have no influence on the behavior of the CML.  This situation
represents the limiting case of $N_{tot}$ independently operating local objects 
at each lattice site. In the general case $0 \le \epsilon \le 1$,  the 
independence of individual cells is 
lost and the lattice behavior is governed by both local and 
global influences. CMLs with a maximal neighborhood, $N \approx N_{tot}$, 
are often denoted as globally coupled maps.    
If the coupling is maximal, $\epsilon\rightarrow 1$,
the behavior of the entire CML is determined by global properties alone (mean field 
approach).

The second term on the rhs in Eq.~(1) contains the states of the neighboring 
map sites at the same time step $n$ at which the first term specifies the state 
of the site whose neighborhood is considered. This type of coupling, assuming
a vanishing transmission time $\Delta t\rightarrow 0$, is sometimes
called ``future coupling'' [Mehta \& Sinha 2000]
since it refers to a situation in which the neighborhood 
states are treated as if they act back from future to present. 
In this sense, we denote this type of coupling as ``non-causal''. 
In order to take a finite transmission time $\Delta t > 0$ into account, 
one can modify the second term in Eq.~(1)
such that $f_n(x_k)$ is replaced by $f_{n-1}(x_k) = x_k$. In this way, past states 
in the neighborhood of a site are considered to act on the present state
of a given site with limited signal speed so that interactions are delayed
rather than instantaneous. Corresponding coupling scenarios, which   
we denote as ``causal'' coupling, have recently been studied by Mehta \& Sinha
[2000], Masoller et al.~[2003], Li et al.~[2004], Atay et al.~[2004], and
Atmanspacher \& Scheingraber [2004] (briefly A \& S).      

Another time scale important for the physical interpretation of Eq.~(1) is
the time interval $\Delta \tau$ assumed for the updating mechanism, i.e.~for
the physical integration of signals from the neighborhood states with the
state considered. If signals between cells are
transmitted much slower than the time scale assumed for the updating mechanism,
$\Delta \tau \ll \Delta t$, the updating can be implemented (almost) instantaneously,
or synchronously. If this is not the case, $\Delta \tau \gtrsim \Delta t$, 
updating must be implemented in an asynchronous way. This entails the 
additional problem of determining a proper updating sequence, which can be 
random or depend on particular features of the situation considered. 

Much of the work on CMLs so far
was based on synchronous 
updating with future, or non-causal, coupling. For asynchronous updating as, for instance, studied by Lumer \& Nicolis 
[1994], it was found that the behavior of CMLs differs strongly from that of CMLs with 
synchronous
updating. Additional results for asynchronous updating were reported by Marcq et 
al.~[1997], Rolf et al.~[1998], Mehta \& Sinha [2000], and A \& S. 
Particularly the last two references studied the
behavior of CMLs under causal rather than non-causal coupling.

\section{Stabilization}

In contrast to the complexity and richness of the phenomenological behavior
of CMLs, investigating their stability properties simplifies the picture and
bases the discussion on more fundamental considerations. The emergence of
global properties from local properties of constitutents of a system are 
particularly interesting examples (cf.~Bishop \&
Atmanspacher [2004]). For this reason, several authors have directed their attention 
to the analysis of the stability in CMLs (e.g., Mackey \& Milton [1995], 
Belykh et al.~[2000], Gade \& Hu [2000], Gelover-Santiago et al.~[2000], 
Mehta \& Sinha [2000], Jost \&Joy [2002], Lin \& Wan [2002], 
Anteneodo et al.~[2003], A \& S).  

As a common feature of the (so far) few studies of asynchronous updating, it has 
been reported that it facilitates the synchronization and stabilization of CMLs
decisively. In particular, Mehta \& Sinha [2000] demonstrated that the dynamics 
at individual lattice cells is strongly synchronized by coupling among cells.
In A \& S we showed that the
local behavior at individual cells can be stabilized under causal coupling
even if it is explicitly unstable. 
In other words, unstable local fixed points become stabilized
as a consequence of their coupling to neighboring unstable fixed points.     


 
In this paper we study the dependence of this stabilization behavior 
upon variation of the degree of causal versus non-causal coupling
for different kinds and sizes of neighborhood and different modes of updating.
For this purpose, we represent the neighborhood term in Eq.~(1) as
a convex combination of causal and non-causal coupling:
\begin{equation}
g_{n+1}(x_{ij}) = (1-\epsilon) f_n(x_{ij}) + \alpha {\epsilon\over N} 
\sum_{k=1}^N f_{n-1}(x_k) + (1-\alpha)  {\epsilon\over N} \sum_{k=1}^N f_{n}(x_k)
\end{equation}
For $\alpha = 0$, the case of full non-causal coupling as in Eq.~(1)
is recovered, while $\alpha = 1$ represents the case of full causal coupling
as studied in A \& S. Note that convex 
combinations of causal and non-causal coupling as in Eq.~(2) represent
the situation of a causal coupling degree $\alpha$ in each neighborhood
site. This is different from a distribution of neighborhood sites with
either full causal or full non-causal coupling even if this would give
rise to the same overall $\alpha$. As an example, a neighborhood consisting
of four sites with $\alpha = 0.5$ leads to behavior which is different
from that obtained for two sites with full causal and two sites with full
non-causal coupling (although this also yields $\alpha = 0.5$ on average).    
  
The following section 2 presents the results of numerical 
simulations of CMLs with synchronous and asynchronous updating, 
for different types and sizes of neighborhoods, for different 
coupling strengths $0 \le \epsilon \le 1$, and for different degrees $\alpha$
of causal coupling. It will be shown how the global stabilization 
of unstable local behavior depends on the degree of causal or
non-causal coupling.  
In section 3, the results for different updating and different
neighborhoods are compared and discussed.
Section 4 summarizes and concludes the paper, and some perspectives will 
be addressed. 

\section{Numerical Results}

In this section, we present results from numerical simulations of two-dimensional 
coupled map lattices according to Eq.~(2). Since the focus of this contribution
is on the stabilization of unstable behavior, we have to work within a parameter
range in which the behavior of the logistic maps at each lattice site is unstable.        

The quadratic function $x \mapsto rx(1-x)$
has two critical points, one at 0 and one at ${r-1\over r}$. 
The stability properties of these critical points are directly related 
to the derivative of the function $x \mapsto rx(1-x)$ at each of them. If the absolute value of the derivative
is smaller (greater) than 1, then the critical point is stable (unstable). 
Hence, the critical point at 0 is a stable fixed point for $r<1$ and unstable
for $r\ge 1$. The critical point at ${r-1\over r}$ is stable for $r<3$ and 
unstable for $r\ge 3$.   

For our investigations, 
we focus on the more interesting unstable fixed point
at ${r-1\over r}$ and use the control parameter $r=4$ to demonstrate the results. 
(As far as the topic of this contribution is concerned, there is no basic
difference in behavior for other values of $r$ as long as $r>3$.)
The corresponding unstable fixed point is located at 0.75.
We study the distribution of state values of a lattice of size $50\times 50$ 
($N_{tot}=2500$ cells with random initial conditions) after a number of iterations which
is large enough that transients have died out, usually after 10000 iteration steps. 

We consider different kinds of neighborhoods according to 
the second term of Eq.~(1). Results for both von Neumann neighborhoods and 
Moore neighborhoods will be presented, each of both order 1 and 2. A von
Neumann neighborhood of order 1 includes the $N=4$ vertically and horizontally
nearest neighbors of a given site. A Moore neighborhood of order 1 includes the
4 diagonal nearest neighbors in addition, hence covering a square of $N=8$ cells 
in total. A von Neumann
neighborhood of order 2 is constructed by a Moore neighborhood of order 1 plus
the vertical and horizontal second neighbors of a given site, hence it consists
of $N=12$ cells in total. A Moore neighborhood of order 2 covers, in addition, all cells
covering a square of side length 5, hence $N=24$ cells in total.     

The behavior of CMLs depends on the way in which
the values at each cell are updated from one to the next iteration step. As two
basic types of updating, we distinguish between synchronous updating, where 
all values are calculated subsequently but updated at once, and asynchronous 
updating, where all values are updated in the sequence in which they are calculated.
For the latter procedure, it is crucial how the sequence is defined. 
In case of asynchronous updating, those cells which are 
already updated affect the behavior of the CML before the update providing the next
iteration step is complete. This does not happen in case of synchronous updating.  

The degree of causal versus non-causal coupling can be controlled by the
parameter $\alpha$ in Eq.~(2). Since updating mechanism and causal coupling introduce 
two time scales for a physically realistic interpretation of Eq.~(2), they are
strictly relevant for the same time step only if $\Delta \tau \approx \Delta t$. 
If this is not the case, we may ``simulate'' the effect of different 
time scales by rescaling 
$\Delta t$ (or considering fractions or multiples of it) so that it
approximates $\Delta \tau$. In this way, it is
possible to study both causal and non-causal coupling in synchronous as well
as asynchronous updating scenarios in a qualitative manner.   
   
\subsection{Synchronous Updating}

For the presentation of our results we use stability diagrams of the same
kind as used in A \& S. 
They are obtained by numerical runs of the CML given by Eq.~(2)
The runs are terminated after 10000 iteration steps for coupling strengths 
$0 \le \epsilon \le 1$ and for a degree $0 \le \alpha \le 1$ of causal coupling. 

\renewcommand{\baselinestretch}{0.85}
\begin{figure}
\begin{center}
\epsfig{figure=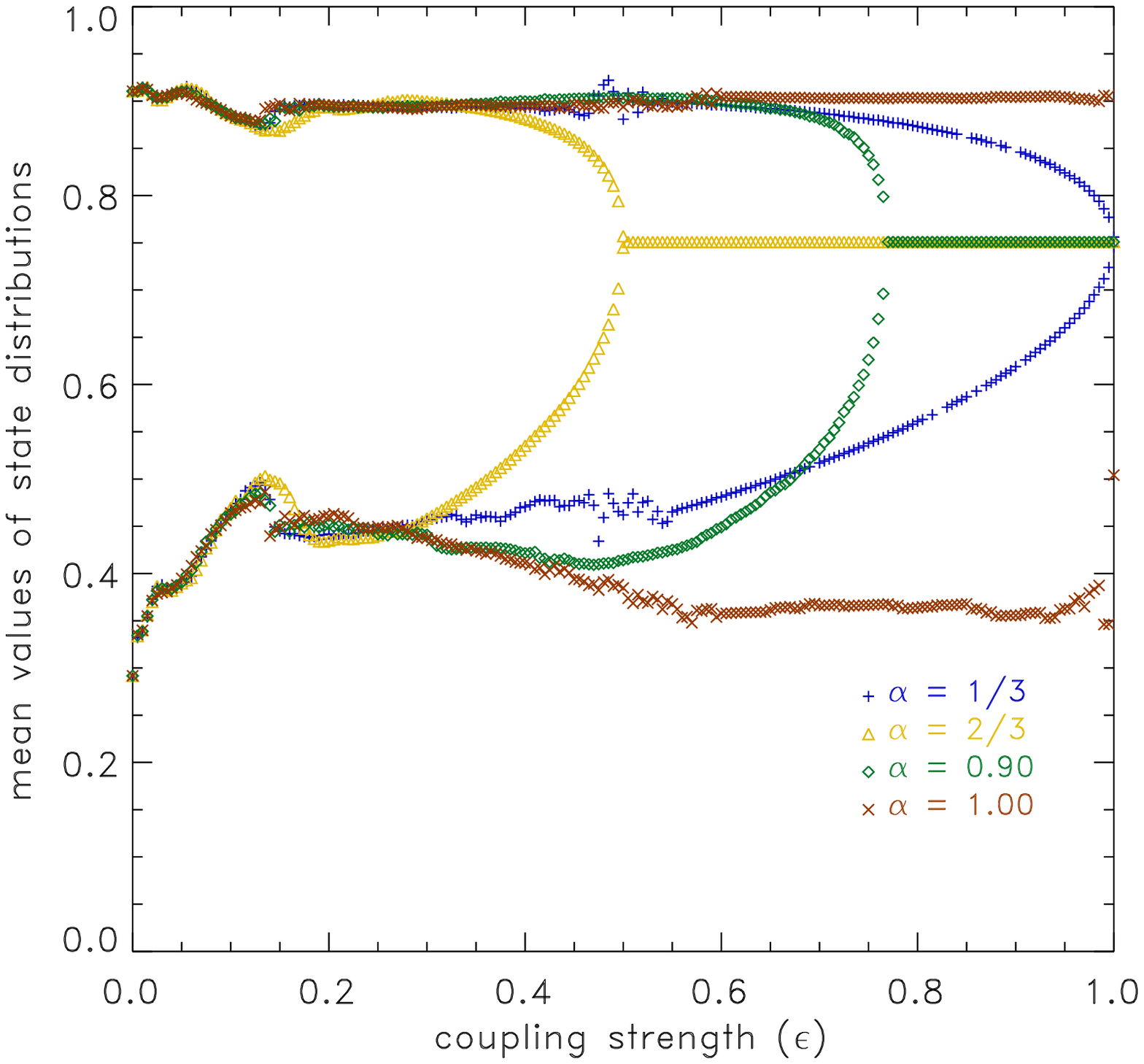,scale=0.6}  
\end{center}
\begin{center}
\epsfig{figure=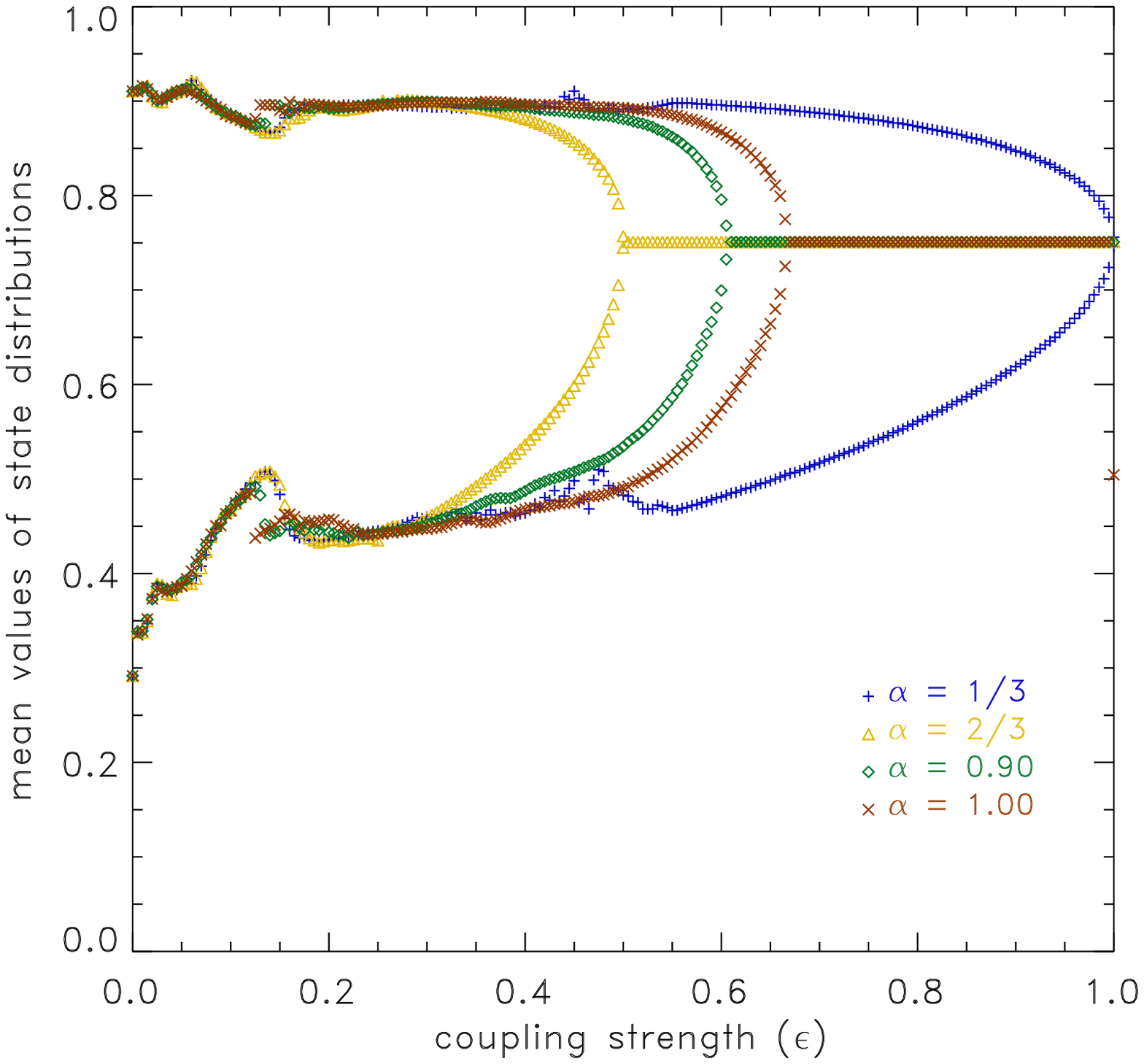,scale=0.6}  
\end{center}
\begin{quote}
{\footnotesize Figure 1: Stability diagram for synchronously updated CMLs 
with (a) von Neumann neighborhood of order 1 (above) and (b) Moore neighborhood 
of order 1 (below).
Mean values of the state distribution right and left of the unstable
fixed point at 0.75, averaged over ten sets of random initial conditions, are plotted
versus the coupling strength $\epsilon$ for selected degrees $\alpha$ of causal 
coupling.  The control parameter of the logistic map is
set at $r=4$.} 
\end{quote}
\end{figure}

\renewcommand{\baselinestretch}{0.85}
\begin{figure}
\epsfig{figure=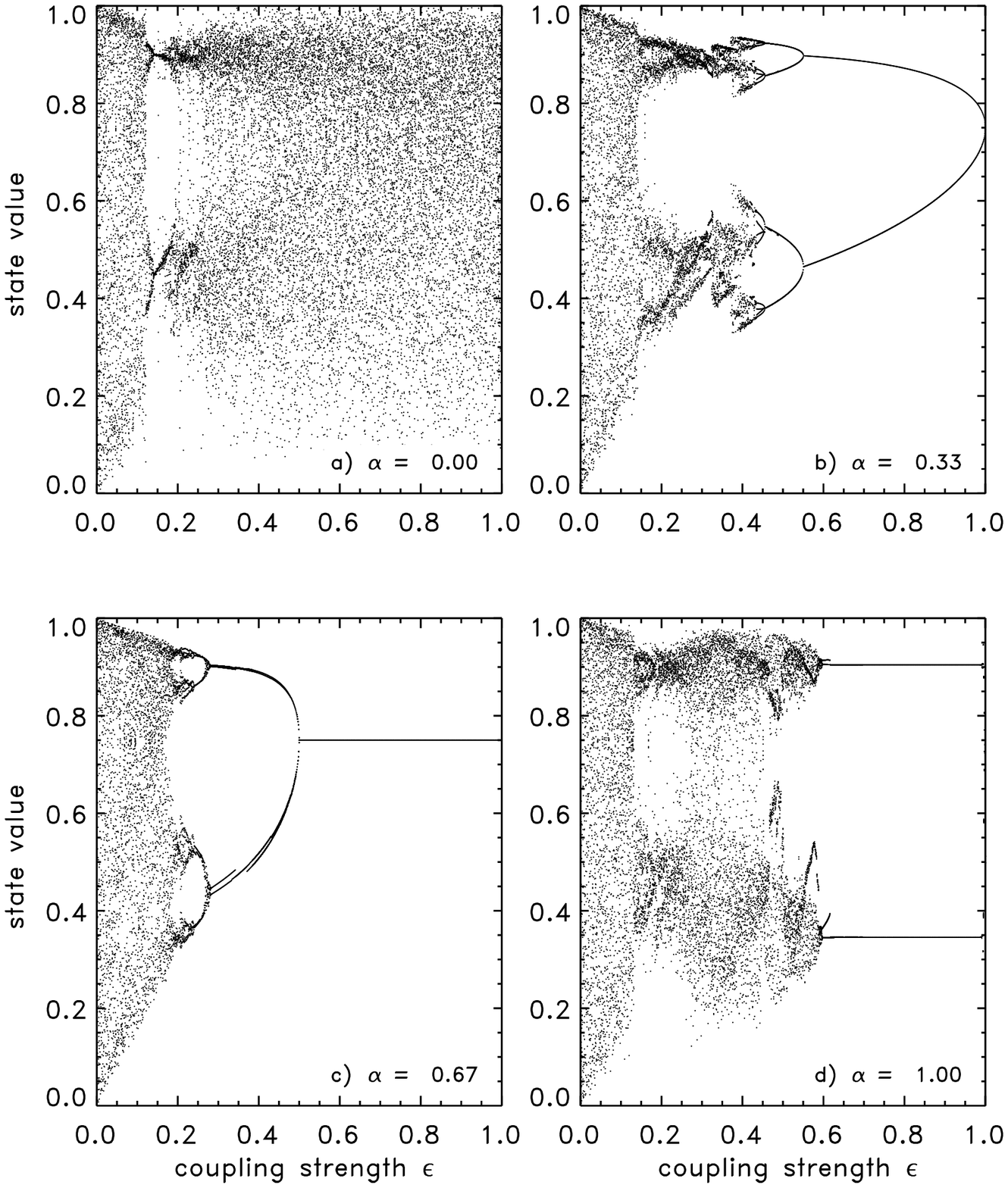,scale=0.85} 
\vspace{0.1cm}
\begin{quote}
{\footnotesize Figure 2: Stability diagram for synchronously updated CMLs 
with von Neumann neighborhood of order 1. 
For selected critical values of $\alpha$, the distribution of states
is plotted versus the coupling strength $\epsilon$. 
The control parameter of the logistic map is set at $r=4$.}     
\end{quote}
\end{figure}

Figure 1a shows stability diagrams 
for a first order von Neumann neighborhood, Fig.~1b shows corresponding diagrams 
for a first order Moore neighborhood. The vertical axis indicates
mean values of the state distributions for $x > x_{unst} = 0.75$ and for
$x < x_{unst} = 0.75$, which are 
averaged over ten different sets of random initial conditions for the CML.
Beyond the critical value of $\epsilon$, at which 
stabilization at the unstable fixed point occurs, the mean value of $x$
is simply the unstable fixed point itself. The stability diagrams are
shown for selected values of $\alpha$.
Error bars are of the size of the symbols. 

For the first order von Neumann neighborhood (Fig.~1a) we recover the
bistable behavior reported in A \& S at full causal coupling $\alpha = 1$. 
However, decreasing (increasing) the degree of causal (non-causal) coupling
leads to stabilization, whose onset value $\epsilon$ varies as a function 
of $\alpha$. Only for quite small causal (high non-causal) coupling
the stabilization disappears and gives room for behavior which is not even
bistable but chaotic.  

Figure 1b shows corresponding results for the first order Moore neighborhood. 
At full causal coupling $\alpha = 1$, the stabilization behavior 
reported in A \& S is reproduced. Moreover, the stabilization onset extends 
to even smaller values of $\epsilon$ when causal (non-causal) coupling is reduced
(enhanced). As for a first order von Neumann neighborhood, stabilization is
completely lost for small causal (high non-causal) coupling.   

For small values of $\epsilon$, the behavior of the stability curves is 
essentially identical. Neighborhhoods of order 2 produce stability diagrams 
similar to the Moore neighborhood of order 1. As shown in A \& S, the critical
value of $\epsilon$ for stabilization onset with full causal coupling decreases 
for an increasing size of the neighborhood.              

The general trend visible in Figs.~1a and 1b is that full causal coupling 
with synchronous updating provides less pronounced stabilization (bistability
in case of the first order von Neumann neighborhood) than 
achieved with small additions of non-causal coupling. If non-causal coupling 
is predominant and causal coupling is quite small, no stabilization occurs
at all. As will be shown and discussed in more detail in section 3, critical 
degrees of causal coupling are $\alpha = 1/3$ and $\alpha = 2/3$. 

Plotting the mean values of the state distribution as a function of $\epsilon$ 
as in Fig.~1 allows us to show stability diagrams for several values of
$\alpha$ in one figure. However, the fine structure of the stability diagram
is only visible if the state distribution itself (rather than its mean values
above and below the unstable fixed point) is plotted as a function
of $\epsilon$. Figure 2 shows such plots for the first order von Neumann
neighborhood (cf.~Fig.~1a) and for $\alpha = 0, 1/3, 2/3, 1$. 
(Higher order neighborhoods provide similar features.)

For $\alpha = 0$, the state distribution is essentially continuous except
in a small regime around $\epsilon \approx 0.2$. At $\alpha = 1/3$, there
is some interesting fine structure in the stability diagram for 
$0.2 \lesssim \epsilon \lesssim 0.5$; for $\epsilon \gtrsim 0.5$ we observe bistable
behavior which becomes globally stable at the unstable fixed point at
maximal coupling, $\epsilon = 1$. 

For $\alpha = 2/3$, the stability diagram for the state distribution yields 
a most regular structure. Stabilization at the unstable fixed point sets in
at $\epsilon = 0.5$, the lowest value at which this is possible. The case of 
$\alpha = 1$ shows much fine structure around $0.6 \lesssim \epsilon \lesssim 0.7$; 
for $\epsilon > 0.75$ we have bistable behavior.            

For both $\alpha=0.33$ and $\alpha=1$, Fig.~2 shows a pronounced fine structure.
It is due to long-time transients which die out only after a number of iterations 
of the order of several $10^5$.   We indicate this surprising result in view
of the fact that transient behavior is, for obvious reasons, not easy to
study in a systematic fashion. Synchronously updated CMLs with
a first order von Neumann neighborhood might open up interesting options 
for such investigations.     

\subsection{Asynchronous Updating}

For an asynchronous updating procedure, the updating sequence needs to be
adapted to the situation which is modeled by the implemented CML. With respect
to the stability properties of the CML, we have demonstrated in A \& S that a 
random selection of the updating
sequence produces results which are representative for linear and value-dependent
asynchronous updating as well.
For random asynchronous updates, Figs.~3a and 3b show two stability diagrams 
of the mean values of states (as in Fig.~1) after 10000 iteration steps for 
a von Neumann neighborhood of order 1 and a Moore neighborhood of order 1 as 
a function of coupling strength $\epsilon$.

\renewcommand{\baselinestretch}{0.85}
\begin{figure}
\begin{center}
\epsfig{figure=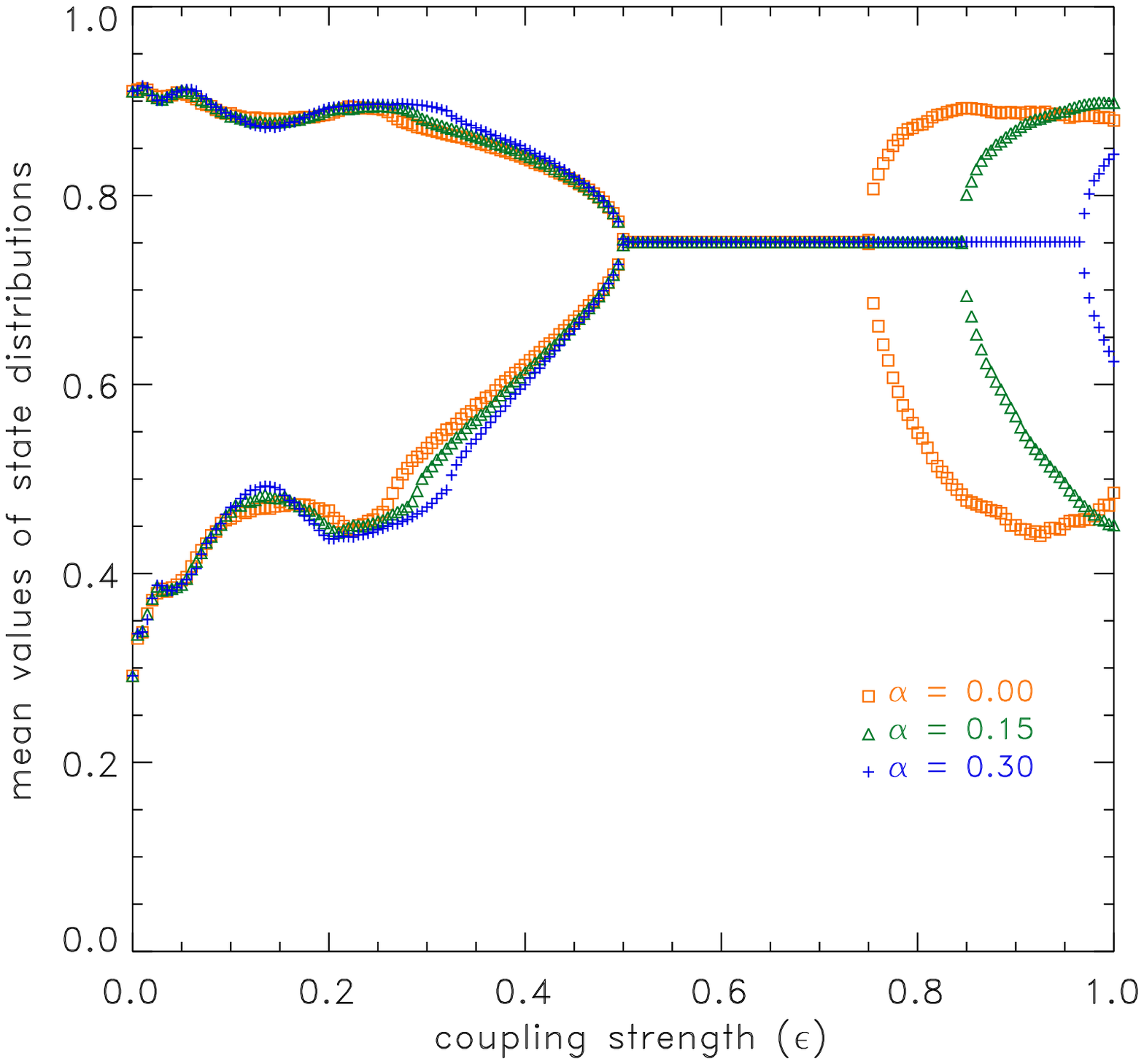,scale=0.6}  
\end{center}
\begin{center}
\epsfig{figure=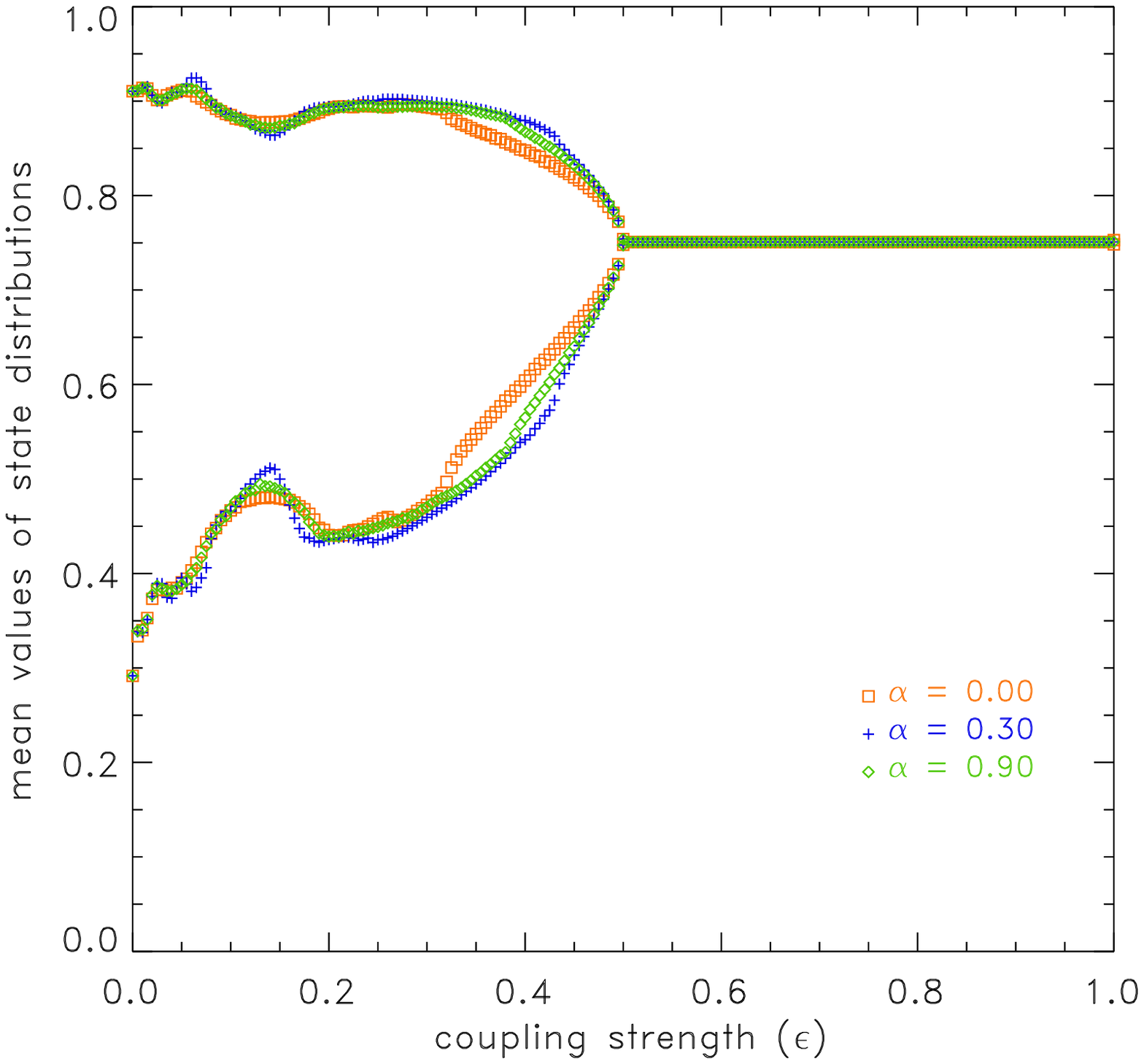,scale=0.6}  
\end{center}
\begin{quote}
{\footnotesize Figure 3: Stability diagram for asynchronously updated CMLs 
with (a) von Neumann neighborhood of order 1 (above) and (b) Moore neighborhood 
of order 1 (below).
Mean values of the state distribution right and left of the unstable
fixed point at 0.75, averaged over ten sets of random initial conditions, are plotted
versus the coupling strength $\epsilon$ for selected degrees $\alpha$ of causal 
coupling.  The control parameter of the logistic map is
set at $r=4$.} 
\end{quote}
\end{figure}

\renewcommand{\baselinestretch}{0.85}
\begin{figure}
\epsfig{figure=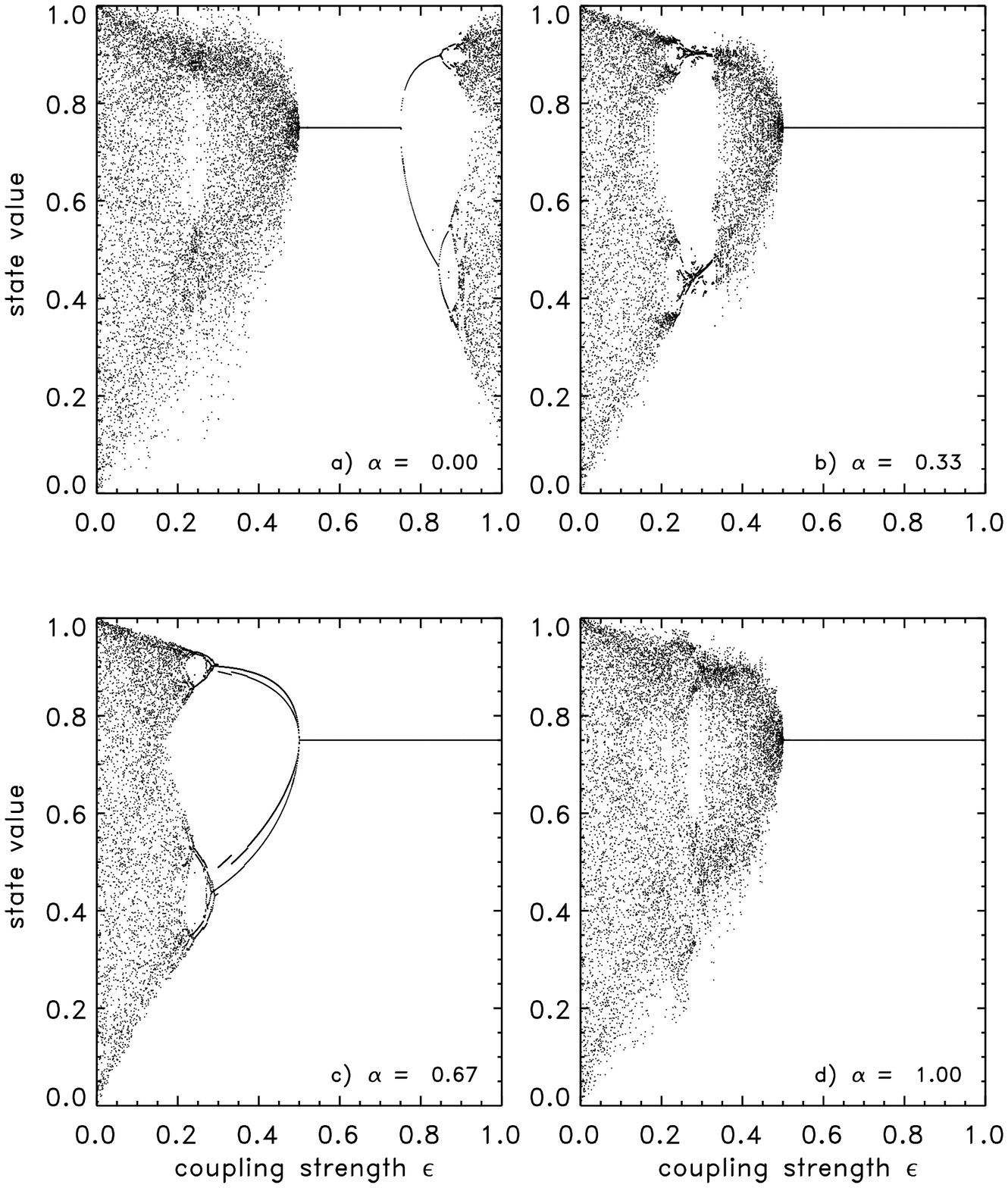,scale=0.85} 
\vspace{0.1cm}
\begin{quote}
{\footnotesize Figure 4: Stability diagram for asynchronously updated CMLs 
with von Neumann neighborhood of order 1. 
For selected critical values of $\alpha$, the distribution of states
is plotted versus the coupling strength $\epsilon$. 
The control parameter of the logistic map is set at $r=4$.}     
\end{quote}
\end{figure}

Independently of $\alpha$, stabilization at the unstable fixed point sets in 
precisely at $\epsilon = 0.5$. For the Moore neighborhood, this stabilization
extends to $\epsilon = 1$ for all values of $\alpha$. However, the von Neumann
neighborhood of first order shows a peculiar bifurcation back into non-stabilized
behavior for higher values of $\epsilon$ if the degree of causal coupling 
$\alpha$ is small enough. The critical value of $\epsilon$ 
for the decay of stabilization in the regime of small causal coupling depends 
on $\alpha$.

In general the stabilization onset of CMLs with asynchronous updating is
robust against the influence of non-causal coupling. Even if non-causal
coupling is maximal and causal coupling vanishes entirely, stabilization
sets in as in the case of full causal coupling studied in A \& S. 
The same happens for higher order neighborhoods, not shown in Fig.~3.

Plotting the state distribution rather than its means above and below
the unstable fixed point, the stability diagram corresponding to Fig.~3a 
(first order von Neumann neighborhood) is shown in Fig.~4 for the same
values of $\alpha$ as in Fig.~2. The case $\alpha = 0$ clearly
shows the bifurcation at $\epsilon = 0.75$. Increasing $\alpha$ to 1/3
has the consequence that the bifurcation point moves to higher coupling 
strengths and disappears precisely at $\alpha =1/3$. 

The situation for $\alpha = 2/3$ is almost indistinguishable from 
the corresponding diagram for synchronous update. Although the
onset of stabilization is at $\epsilon = 0.5$ independently of $\alpha$,
the stability diagram in Fig.~4 shows clearly that most regularity
is obtained at $\alpha =2/3$. For $\alpha = 1$ we see more or less
the same features as for $\alpha = 0$ in the range of $\epsilon < 0.5$.

One peculiar feature for complete non-causal coupling $\alpha = 0$ 
occurs if one focuses on the bifurcation at  $\epsilon = 0.75$ for different 
numbers of iteration steps. The feature shown in Fig.~2 
(upper left) is obtained after transients have definitely died out. 
Reducing the number of iteration steps does not lead to less 
structure, however,  but implies an increasingly complicated,
yet regular fine structure. Similar to particular situations for
synchronous updating, asynchronously updated CMLs with
a first order von Neumann neighborhood for a low degree of causal
coupling are interesting candidates for such investigations.


\vfill \eject

\section{Discussion}  

Figures 1--4 showed how synchronously and asynchronously updated
CMLs with different neighborhoods are stabilized and destabilized, 
respectively, as a function of coupling strength. The stability diagrams
represent different curves for different (selected) degrees of causal 
coupling $\alpha$ and non-causal coupling $(1-\alpha)$, respectively. 
Since the stability behavior of the CMLs depends strongly on $\alpha$,  
it is worthwhile to consider the full range between causal and non-causal
coupling in detail.  

For the purpose of a convenient illustration, we determine those critical
value(s) of $\epsilon$ at which either stabilization sets in or decays. 
These values $\epsilon_{crit}$ determine either an inverse bifurcation or a
bifurcation in Figs.~1--4. In Fig.~5, they are plotted as a function
of $\alpha$. Full causal coupling as investigated in A \& S is represented
by the values of $\epsilon_{crit}$ at $\alpha = 1$. 

\begin{figure}[h]
\begin{center}
\vskip 0.2cm
\epsfig{figure=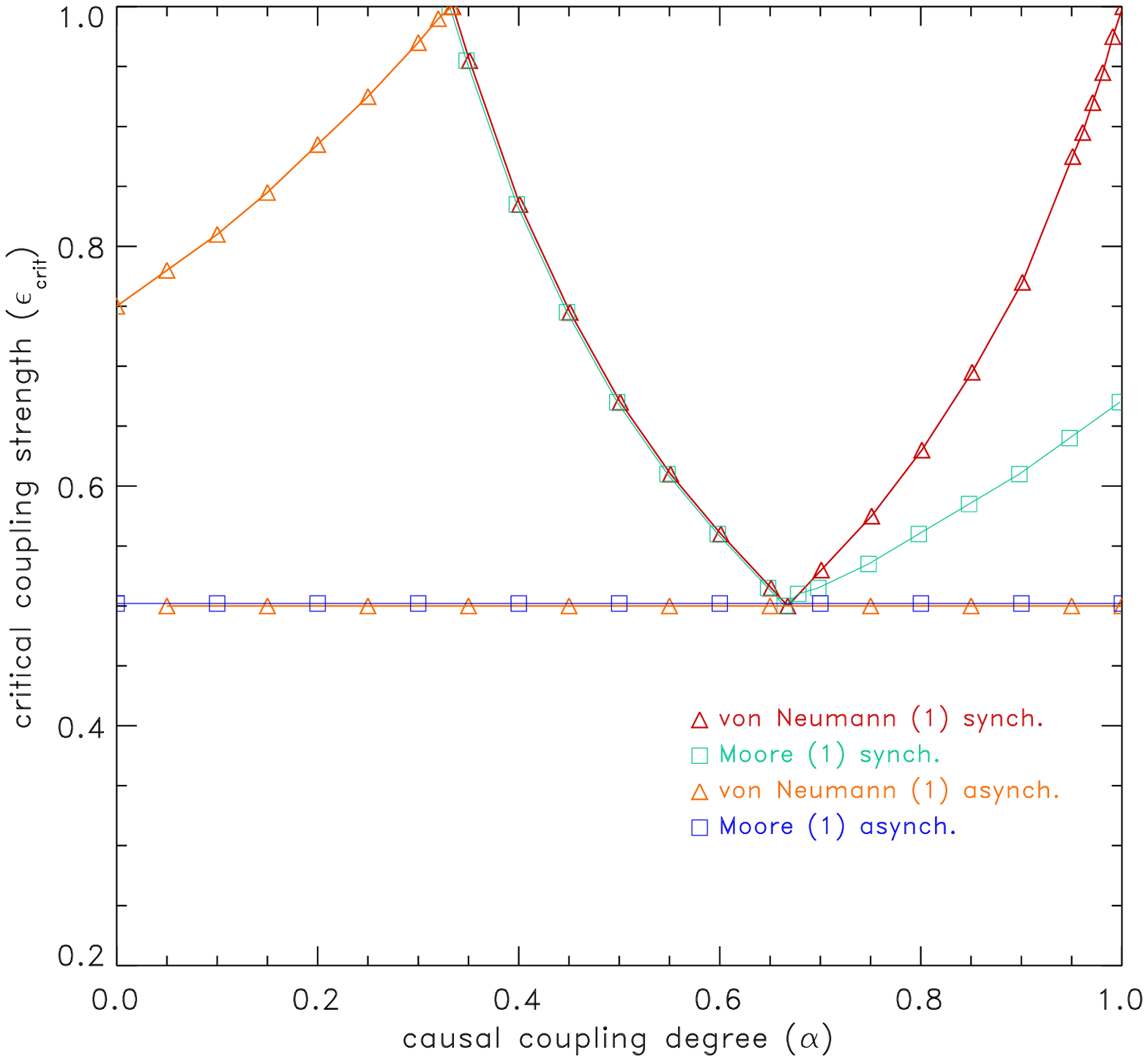,scale=0.7}
\end{center}
\begin{quote} {\footnotesize Figure 5: Critical coupling strength $\epsilon_{crit}$ 
for stabilization onset as a funtion of the degree $\alpha$ of causal coupling.
Different symbols refer to different updating procedures and different
neighborhoods as explained in the figure.}    
\end{quote}
\end{figure}

Let us first discuss the case of asynchronous updating. For both Moore and
von Neumann neighborhoods of order 1, $\epsilon_{crit} = 0.5$ marks the
stabilization onset of the CML. As derived in detail in A \& S, this
corresponds to the lower bound for the coupling strength $\epsilon$ beyond 
which stabilization becomes possible at all. The horizontal line for 
$\epsilon_{crit} = 0.5$ in Fig.~5 shows that this is independent of the
degree of causal coupling. 

As an interesting additional feature, which does not occur for the Moore 
neighborhood of order 1, observe the destabilizing bifurcation in Figs.~2a 
and 3 for 
$\alpha < 1/3$ at $\epsilon_{crit} > 0.75$ in case of the von Neumann
neighborhood of order 1. It shows that strong non-causal coupling leads
to a partial loss of stability in this special case. The stability of
higher order types of neighborhood (including global coupling) for asynchronous 
updating is robust
against the influence of different degrees of (non-) causal coupling.        

This is quite different for the situation of synchronous update. As can be
seen in Fig.~5, there is no stabilization at all for $\alpha < 1/3$. 
In the regime $1/3 < \alpha < 2/3$, the value of $\epsilon_{crit}$ characterizing
stabilization onset is a monotonically 
decreasing function of $\alpha$.  
At $\alpha = 2/3$, the value of $\epsilon_{crit}$ reaches its lower bound
of 0.5, as for asynchronous updating. This behavior is identical for all 
neighborhoods, including global coupling.

In the regime $2/3 < \alpha < 1$, 
$\epsilon_{crit}$ increases monotonically for increasing $\alpha$. 
This increase, however, is different for different neighborhoods. 
The von Neumann neighborhood of order 1 provides the limit of
$\epsilon_{crit} \rightarrow 1$ for $\alpha\rightarrow 1$, again
showing the peculiarity of this type of neighborhood. For an increasing
neighborhood size, $\epsilon_{crit}$ becomes flatter as a function of
$\alpha$ and approaches the limit of a constant stabilization
onset at $\epsilon_{crit} = 0.5$ for global coupling.

\section{Summary and Conclusions}

The behavior of coupled map lattices (CMLs) depends on a number of conditions,
whose variation leads to an immense richness of phenomenological features
which are difficult to classify. However, the emergence or decay of such features 
relies on basic stability properties of the CMLs which allow us to reduce the 
phenomenological complexity and, thus, provide a simpler picture. In particular,
the global stabilization of locally unstable behavior is illustrative in this 
respect.

The crucial conditions, on which such global stabilization depends, are
the form of the individual maps, the size and type of their neighborhoods, 
the strength $\epsilon$ of the coupling between individual maps
and their neighborhood, and the time scales of the updating mechanism 
(synchronous or asynchronous) and of the interaction among maps. 

Following up on earlier investigations of CMLs using the logistic map at
$r = 4$ with synchronous and asynchronous updating for low order von Neumann
and Moore neighborhoods, this paper presents numerical results for the 
stabilization of CMLs as a function of different time
delays between individual maps and their neighborhood. The limiting cases
for this delay are instantaneous interaction on the one hand 
(non-causal coupling) and finite-time interaction on the other (causal 
coupling).  

Studying the full range between pure causal and pure non-causal coupling
in terms of a convex combination of the two, we found that the degree 
$\alpha$ of causal coupling in a CML and its stabilization behavior are 
related in surprising ways.
\begin{itemize} 
\item For {\it asynchronous updating}, the critical coupling strength 
$\epsilon_{crit} = 0.5$ for stabilization onset is in general robust 
against variations of both the degree of causal coupling and the type and
size of neighborhood. The von Neumann neighborhood of order 1 shows an
additional destabilization in the regime of small causal coupling, which is 
not observed for all other neighborhoods.
\item For {\it synchronous updating}, there is no stabilization at all for
small causal coupling $\alpha < 1/3$. In the regime $1/3 < \alpha < 1$,
the influence of causal coupling induces stabilization at different critical
coupling strengths. For global coupling with a causal degree $\alpha > 2/3$, 
the stabilization onset coincides with that of asynchronous updating.
\end{itemize}

The reasons for the precise behavior of the stabilization onset as a function 
of causal coupling remain to be understood in detail. We expect that such
an understanding will provide insight into the relation between causality and
stability in general. Since CMLs can be considered as discretized partial 
differential equations, our approach might be a potential candidate for studying superpositions of advanced and retarded solutions of time-reversal invariant 
signal transmission. Our numerical results suggest the significance of
stability criteria for distinguishing retarded, i.e.~causal, solutions  
under conditions corresponding to synchronous updating.

\section{References}

\begin{description}


\item Anteneodo, C., de S.~Pinto, S.E., Batista, A.M., \& Viana, R.L.~[2003]:
``Analytical results for coupled-map lattices with long-range interactions'',
{\it Physical Review E} {\bf 68}, 045202(R). 

\item Atay, F., Jost, J., \& Wende, A.~[2004]: ``Delays, connection topology,
and synchronization of coupled chaotic maps'', lanl preprints cond-mat/0312177.



\item Atmanspacher, H., \& Scheingraber, H.~[2004]: ``Inherent global stabilization 
of unstable local behavior in coupled map lattices'', submitted.

\item Belykh, V., Belykh, I., Komrakov, N., \& Mosekilde, E.~[2000]: ``Invariant
manifolds and cluster synchronization in a family of locally coupled map lattices'',
{\it Discrete Dynamics in Nature and Society} {\bf 4}, 245--256.

\item Bishop, R., \& Atmanspacher, H.~[2004]: ``Contextual emergence in the
description of properties'', submitted.

\item Gade, P.M., \& Hu, C.-K.~[2000]: ``Synchronous chaos in coupled map lattices
with small-world interactions'', {\it Physical Review E} {\bf 62}, 6409--6413.

\item Gelover-Santiago, A.L., Lima, R., \& Martinez-Mekler, G.~[2000]:
``Synchronization and cluster periodic solutions in globally coupled maps'',
{\it Physica A} {\bf 283}, 131--135.




\item Jost, J., \& Joy, M.P.~[2002]: ``Spectral properties and synchronization
in coupled map lattices'', {\it Physical Review E} {\bf 65}, 016201.

\item Kaneko, K., ed.~[1993]: {\it Theory and Applications of Coupled Map 
Lattices}, Wiley, New York.

\item Kaneko, K., \& Tsuda, I.~[2000]: {\it Complex Systems: Chaos and Beyond},
Springer, Berlin.



\item Li, C., Li, S., Liao, X., \& Yu, J.~[2004]: ``Synchronization in coupled
map lattices with small-world delayed interactions'', 
{\it Physica A} {\bf 335}, 365--370.

\item Lin, W.-W., \& Wang, Y.-Q.~[2002]: ``Chaotic synchronization in coupled map
lattices with periodic boundary conditions'', {\it SIAM Journal of Applied
Dynamical Systems} {\bf 1}, 175--189.

\item Lumer, E.D., \& Nicolis, G.~[1994]: ``Synchronous versus asynchronous dynamics
in spatially distributed systems'', {\it Physica D} {\bf 71}, 440--452.  

\item Mackey, M., \& Milton, J.~[1995]: ``Asymptotic stability of densities in coupled 
map lattices'', {\it Physica D} {\bf 80}, 1--17.

\item Marcq, P., Chat\'e, H., \& Manneville, P.~[1997]: ``Universality in Ising-like
phase transitions of lattices of coupled chaotic maps'', {\it Physical Review E}
{\bf 55}, 2606--2627. 

\item Masoller, C., Marti, A.C., \& Zanette, D.H.~[2003]: ``Synchronization in an
array of globally coupled maps with delayed interactions'', {\it Physica A}
{\bf 325}, 186--191.

\item Mehta, M., \& Sinha, S.~[2000]: ``Asynchronous updating of coupled maps leads
to synchronization'', {\it CHAOS} {\bf 10}, 350--358. 


\item Rolf, J., Bohr, T., \& Jensen, M.H.~[1998]: ``Directed percolation universality
in asynchronous evolution of spatiotemporal intermittency'', {\it Physical Review E}
{\bf 57}, R2503--R2506. 


\item Turing, A.~[1952]: ``The chemical basis of morphogenesis'', {\it Transactions
of the Royal Society London, Series B} {\bf 237}, 37--72.

\end{description}
\end{document}